\documentclass[aps,prl,twocolumn,groupedaddress]{revtex4}
\usepackage{amsmath}
\usepackage{amssymb}
\usepackage{graphicx}

\begin{document}

\title{Minimum Electrical and Thermal Conductivity of Graphene:\\ A Quasiclassical Approach}

\author{Maxim Trushin and John Schliemann}

\affiliation{Institute for Theoretical Physics, University of Regensburg,
D-93040 Regensburg, Germany}

\date{\today}

\begin{abstract}
We investigate the minimum conductivity of graphene within a quasiclassical
approach taking into account electron-hole coherence effects which
stem from the chiral nature of low energy excitations.
Relying on an analytical solution
of the kinetic equation in the electron-hole coherent and incoherent cases we 
study
both the electrical and thermal conductivity whose relation fullfills 
Wiedemann--Franz law.
We found that the most of the previous findings based on the Boltzmann 
equation are restricted
to only high mobility samples where electron-hole coherence effects
are not sufficient.

%The minimum electrical conductivity is governed
%by the conductivity quantum within non-universal
%prefactor depending on the scattering potential.
\end{abstract}

%\showpacs
\maketitle

{\em Introduction.} Single graphite layers (graphene) have been
found in the free state only recently \cite{Science2004novoselov},
and their transport properties
have immediately attracted much attention from both experimental
\cite{Nature2005novoselov,Nature2005zhang,PRL2006morozov,Science2007miao,cond-mat2007cho,cond-mat2007chen,cond-mat2007tan}
and theoretical
\cite{PRL2007nomura,kubo,EPJB2006katsnelson,landauer,boltzmann,PRL2006altland,weak,Nature2006katsnelson}
investigators.
The reason of such an explosive interest
is a number of very unusual transport properties including
(i) a non-vanishing electrical conductivity even at zero carrier concentration
(minimum conductivity phenomena),
and (ii) independence of this minimum conductivity of temperature.
Besides these unconventional transport properties
many related phenomena have been studied in graphene such as
weak localisation 
\cite{weak},
the Klein paradox \cite{Nature2006katsnelson} etc.
The remarkable electronic properties of graphene are usually attributed to
the particular spectrum of excitations \cite{PR1947wallace}
which consists of two conical bands
and is described by a two-dimensional analog of the relativistic Dirac equation.
For a review concerning the history, fabrication,
fundamental properties, and future applications of graphene
we refer to the recent article \cite{Nature2007geim}.

To investigate the transport properties (i) and (ii) of graphene
several different approaches have been applied including the Kubo formalism 
\cite{PRL2007nomura,kubo,EPJB2006katsnelson},
direct calculations of the transmission probability for ballistic samples
\cite{EPJB2006katsnelson,landauer},
and Boltzmann equation \cite{PRL2007nomura,boltzmann}.
The latter approach looks at the first sight inapplicable for investigation
of the minimum conductivity since the quasiclassical description is expected to fail at low Fermi energies $E_F$
as soon as $E_F\tau(E_F)$ becomes comparable with $\hbar$. The carrier momentum relaxation time $\tau$,
on the other hand, diverges at $E_F=0$ \cite{PRL2007nomura} for short-range scatterers studied here as well.
Thus, the product $E_F\tau(E_F)$ depends at $E_F\rightarrow 0$ on the scattering parameters rather than
the carrier concentration and can acquire values much lager than $\hbar$ even at zero doping.
In this Letter we solve the kinetic equation for
Dirac fermions including off-diagonal elements of the distribution function in the helicity basis
which are strongly connected with the electron-hole coherence or Zitterbewegung
effects studied recently in graphene by Auslender and Katsnelson \cite{cond-mat2007auslender}.
We find that the off-diagonal elements essentially contribute
to the minimum conductivity in low mobility samples,
and, thus, the conclusions obtained in \cite{boltzmann}
are restricted to only quite perfect graphene sheets.
To discuss our findings we adduce both the electron-hole coherent and conventional
solutions of the kinetic equation. As an application the electrical and thermal conductivity is
calculated.

{\em Preliminaries.}
The carriers in the $\pi$-system of graphene near half filling can be 
described by the Dirac Hamiltonian
\begin{equation}
\label{main_ham}
H=\hbar v_0 (\sigma_x k_x+\sigma_y k_y),
\end{equation}
where
$v_0\approx 10^6 \mathrm{ms^{-1}}$ is the effective ``speed of light'',
$\sigma_{x,y}$ are the Pauli matrices, and ${\mathbf k}$ is 
the two-component particle momentum.
The eigenstates of (\ref{main_ham}) have the form
\begin{equation}
\label{psi}
\Psi_{\mathbf{k}\pm}(x,y)=\frac{1}{\sqrt{2}}{\mathrm e}^{ik_x x+ik_y y}\left(\begin{array}{c}
1 \\ 
\pm {\mathrm e}^{i\theta}
\end{array} \right),
\end{equation}
where $\tan\theta=k_y/k_x$, and the energy spectrum reads
$E_{k\pm}=\pm \hbar v_0 k$. The velocity matrix in the basis (\ref{psi}) is
\begin{equation}
\label{v}
\frac{{\mathbf v}}{v_0} ={\mathbf e_x}\left(
\begin{array}{cc}
\cos\theta  & -i\sin\theta \\ 
i\sin\theta & -\cos\theta
\end{array}\right) + {\mathbf e_y} \left(
\begin{array}{cc}
\sin\theta & i\cos\theta \\
-i\cos\theta & -\sin\theta
\end{array}\right).
\end{equation}

Let us first consider the Boltzmann equation for 
charge carriers in the presence of impurity scatterers
described by the effective potential
\begin{equation}
\label{coulomb}
V(\mathbf{r})=\frac{qeZ}{r}{\mathrm e}^{-r/R}\,
\end{equation}
where $eZ$, $q$ are the impurity atom and carrier electrostatic charge
respectively, and $R$ is the screening radius.
The potential (\ref{coulomb}) differs from its conventional short-range
$\delta$-function approximation \cite{PRL2007nomura}
by the additional fitting parameter $R$.
As we shall see below, the minimum conductivity value is governed
by {\em both} the impurity concentration and screening radius whereas the carrier mobility turns out to be $R$-independent.
It seems to be necessary to introduce such a parameter in order to explain
the experimental picture \cite{cond-mat2007tan}
where two samples made out of the same graphene flake (i. e. having equal mobility)
demonstrate essentially different minimum conductivity values.
This difference is attributed to the screening parameter $R$ in our model.

{\em Electron-hole incoherent solution.}
In linear order in the homogeneous electric field  ${\mathbf E}$ the Boltzmann equation reads
\begin{equation}
\label{master2}
\left(\frac{df_\kappa}{dt}\right)^\mathrm{coll}
=-q{\mathbf E} \mathbf{v}_\kappa \left[-\frac{\partial f^0(E_{k \kappa})}{\partial
E_{k \kappa} }\right]\,,
\end{equation}
where we have divided the distribution function
$f_\kappa({\mathbf k})=f^0(E_{k \kappa})+f^1_\kappa({\mathbf k})$ into an equilibrium contribution
$f^0(E_{k \kappa})$ and a nonequilibrium part $f^1_\kappa({\mathbf k})$, and 
$\mathbf{v}_\kappa$ (with $\kappa \in \{\pm\}$)
are the diagonal elements of the velocity operator in the helicity basis, 
cf. Eq.~(\ref{v}).
Assuming elastic scattering fulfilling the microreversibility condition, 
the collision term can be written as
\begin{equation}
\label{st}
\left(\frac{df_\kappa}{dt}\right)^\mathrm{coll}=\sum\limits_{\kappa'}\int\frac{d^2k'}{\pi^2}
\left\{w(\mathbf{k}\kappa,\mathbf{k'}\kappa')[f_{\kappa'}(\mathbf{k'})
-f_\kappa(\mathbf{k})]\right\},
\end{equation}
where the  scattering probability $w(\mathbf{k}\kappa,\mathbf{k'}\kappa')$
can be easily found from Fermi's 
golden rule,
\begin{eqnarray}
w({\mathbf k}\kappa,{\mathbf k}'\kappa') & = &
\frac{4 \pi R^2 V_0^2}{\hbar}\delta(E_{k \kappa}-E_{k' \kappa'})\times \\ 
\nonumber
&& \frac{1+\kappa \kappa'\cos(\theta'-\theta)}{1+R^2\left[k^2+k'^2-2kk'\cos(\theta'-\theta)\right]}\,.
\label{w1}
\end{eqnarray}
Here $V_0=\pi qeZ \sqrt{N}$ is an effective scattering potential
with $N$ being the impurity concentration, and the fourfold 
(valley and spin) degeneracy
is taken into account by the factor $4$ in the integrand.
The exact analytical solution of Eq.~(\ref{master2}) can be given in terms of 
the nonequilibrium part of the distribution function
$f^1_\kappa=q\mathbf{E}\mathbf{v}_\kappa \tau(k)
\left[-\partial f^0(E_{k \kappa})/\partial E_{k \kappa} \right]$
with $\tau(k)$ given by
\begin{eqnarray}
\label{tau}
\tau(k)&=& \frac{1}{k} \frac{\hbar^2 v_0}{4R^2 V_0^2}\frac{2R^4k^4}{1+2R^2k^2-\sqrt{1+4R^2k^2}}\\
\label{tau0}
&\approx&\frac{1}{k}\frac{\hbar^2 v_0}{4 R^2 V_0^2},
\quad Rk\ll 1.
\end{eqnarray}

{\em Electron-hole coherent solution.}
So far we have neglected the off-diagonal elements of
the distribution function, an valid approximation if the decoherence 
of the single-carrier state is a fast process compared to its relaxation
described by the above equations.
In general, a particle described
by the Hamiltonian (\ref{main_ham}) can not only be in one of the states
$\Psi_{{\mathbf k}+}$ or $\Psi_{{\mathbf k}-}$ but in an arbitrary superposition of them.
Therefore, generalising the above considerations, the distribution function
is $2\times 2$ nondiagonal matrix $\hat{f}({\mathbf k})$,
and the kinetic equation contains the commutator $\frac{i}{\hbar}\left[H,\hat{f}({\mathbf k})\right]$,
which drops out if only the diagonal elements of $\hat{f}({\mathbf k})$
(with respect to the helicity basis) are retained.
In the linear response regime the kinetic equation explicitly reads
\begin{eqnarray}
\nonumber && 
\left(\frac{d\hat{f}}{dt}\right)^\mathrm{coll}=
\frac{i}{\hbar}\left(
\begin{array}{cc}
0 & f_{12}\left(E_{k+}-E_{-}\right) \\  f_{12}\left(E_{k+}-E_{-}\right) & 0
\end{array}\right) \\
\nonumber &&
+q{\mathbf E}\left(
\begin{array}{cc}
 -\mathbf{v}_{11} \left[-\frac{\partial f^0(E_{k +})}{\partial E_{k +} }\right] &
\frac{\mathbf{v}_{12}}{2E_{k+}}\left( f^0_{E_{k -}} - f^0_{E_{k +}} \right) \\ 
\frac{\mathbf{v}_{21}}{2E_{k-}}\left( f^0_{E_{k +}} - f^0_{E_{k -}} \right)
& -\mathbf{v}_{22} \left[-\frac{\partial f^0(E_{k -})}{\partial E_{k -} }\right]
\end{array}\right)\\
\label{master3}
\end{eqnarray}
with the collision term given by
the generalised expression \cite{JETP1984dyakonov}
\begin{eqnarray}
\nonumber
&&
\left(\frac{d\hat{f}}{dt}\right)^\mathrm{coll}_{\kappa\kappa_1}=
\int\frac{d^2 k'}{\pi^2}\sum\limits_{\kappa',\kappa'_1}
\{[\delta(E_{k' \kappa'}-E_{k \kappa}) \\
\nonumber &&
+\delta(E_{k'\kappa'_1}-E_{k\kappa})]
K^{\kappa\kappa_1}_{\kappa' \kappa'_1}f_{\kappa' \kappa'_1}(k')
-\delta(E_{k\kappa'}-E_{k' \kappa'_1}) \\
&&
\times [K^{\kappa\kappa'}_{\kappa'_1 \kappa'_1}f_{\kappa' \kappa_1}(k)+K^{\kappa'\kappa_1}_{\kappa'_1 \kappa'_1}f_{\kappa'_1 \kappa'_1}(k)]\},
\end{eqnarray}
and $K_{\kappa'\kappa'_1}^{\kappa\kappa_1}$ being
\begin{eqnarray}
\nonumber &&
K_{\kappa'\kappa'_1}^{\kappa\kappa_1} =
(\pi R^2 V_0^2/\hbar) \\
&&
\times\frac{1+\kappa\kappa'\kappa_1\kappa'_1+\kappa\kappa'{\mathrm e}^{i(\theta'-\theta)}
+\kappa_1\kappa'_1{\mathrm e}^{-i(\theta'-\theta)}}{1+R^2\left[k^2+k'^2-2kk'\cos(\theta'-\theta)\right]}.
\label{k1}
\end{eqnarray}
Note, that the collision term includes off-diagonal elements of $\hat{f}$ and in that way significantly enhances
the complexity of the kinetic equation.
However, by somewhat more tedious calculations one can
again construct an analytical solution to
Eq.~(\ref{master3}) with nonequilibrium terms given by
\begin{eqnarray}
\label{f11} \nonumber && 
f_{11}^1=q\mathbf{E}\mathbf{v}_{11} \tau(k)\left\{\left(1+\frac{1}{2\alpha}\right)
\left[-\frac{\partial f^0(E_{k+})}{\partial E_{k+} }\right]\right.\\
&&
\left.
+ \frac{1}{2\alpha} \left[-\frac{\partial f^0(E_{k-})}{\partial E_{k-} }\right]
+\frac{1}{2\alpha E_{k+}}\left( f^0_{E_{k +}} - f^0_{E_{k -}} \right)\right\} \\
\label{f12} \nonumber &&
f_{12}^1=\frac{q\mathbf{E}\mathbf{v}_{12} \tau(k)\left(\frac{1}{2}+\frac{1}{2\alpha}\right)}{1+2iE_{k+}\tau(k)/\hbar}
\left\{\frac{1}{E_{k+}}\left( f^0_{E_{k +}} - f^0_{E_{k -}} \right)
\right.\\
&&
\left.
+\left[-\frac{\partial f^0(E_{k+})}{\partial E_{k+} }-\frac{\partial f^0(E_{k-})}{\partial E_{k-} }\right]\right\},
\end{eqnarray}
and $f_{22}^1$, $f_{21}^1$ can be obtained from Eqs.~(\ref{f11}--\ref{f12})
just exchanging the indices belong to $E_k$ and $\mathbf{v}$ accordingly.
Here we have introduced the novel electron-hole incoherence parameter $\alpha=4E_{k+}^2\tau^2(k)/\hbar^2$.
In order to simplify the solution we use
$\tau(k)$ given by Eq.~(\ref{tau0}), thus,
$\alpha=\hbar^4 v_0^4/4R^4V_0^4$ is independent on $k$.
In the limit case of weak scattering ($\alpha\gg 1$)
the diagonal elements $f_{11}^1$ and $f_{22}^1$
are the same as in the electron-hole incoherent case and
given by $f_\kappa^1$, $\kappa \in \{\pm\}$, whereas
the off-diagonal elements $f_{12}^1$ and $f_{21}^1$ are real and,
as we shall see from Eq.~(\ref{current}),
do not contribute to the current.

{\em Electrical conductivity.} The electrical current reads
\begin{equation}
\label{current}
\mathbf{j}=4q\int\frac{d^2 k}{(2\pi)^2}\left(\mathbf{v_{11}}f_{11}+\mathbf{v_{22}}f_{22}
+2\Im \mathbf{v_{12}}\Im f_{12} \right),
\end{equation}
where the factor $4$ is due to the fourfold degeneracy.

Let us first concentrate on the electron-hole incoherent case when $\alpha\gg 1$.
Utilising Eq.~(\ref{tau0}) we find from Eq.~(\ref{current}) for the conductivity
\begin{equation}
\label{cond0}
\sigma\equiv\sigma_0=\frac{q^2 \hbar v_0^2 }{4 \pi R^2 V_0^2}\equiv\frac{q^2 \sqrt{\alpha}}{h},
\end{equation}
which, in particular, does neither depend on temperature nor on 
the carrier concentration. Thus, the most striking features of the 
experimental findings
are reproduced: (i) the conductivity is not zero even at zero carrier 
concentration,
(ii) this minimum conductivity does not depend on temperature.
Note, that the above result is {\em not universal}, i.e. the 
minimum conductivity at zero Fermi energy can change from sample to sample
in accordance with recent experimental reports \cite{cond-mat2007tan,cond-mat2007chen,novoselov}.
Moreover $\sqrt{\alpha}> 4$ (i.e. $1/\alpha< 0.0625$) for the vast majority of samples \cite{cond-mat2007tan,cond-mat2007chen}.

\begin{figure}
\includegraphics[width=\columnwidth]{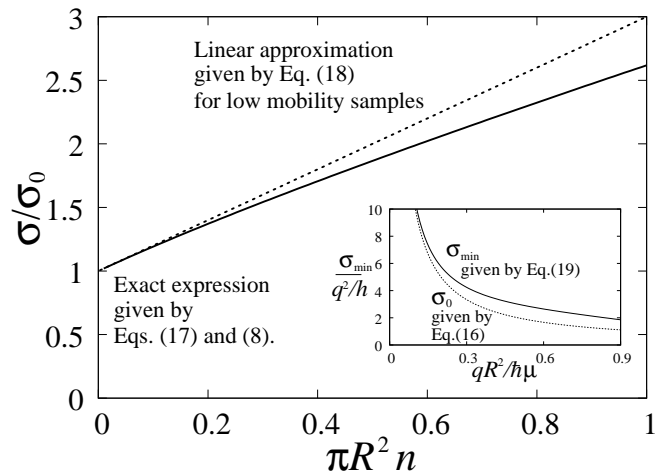}
\caption{\label{fig1} Conductivity dependence on the carrier concentration,
cf. Ref.~\cite{cond-mat2007tan}. Inset: the minimum conductivity vs. inverse
mobility, cf. Ref.~\cite{cond-mat2007chen}. In order to preserve the validity of the quasiclassical
description the parameters are chosen so that $\alpha > 1$. The electron-hole
coherence correction is nevertheless clearly seen.}
\end{figure}

Let us take into account higher-order terms in $\tau(Rk)$.
Then instead of Eq.~(\ref{cond0}) we have at zero temperature
\begin{eqnarray}
\label{conc1}
\sigma & = &\frac{q^2}{\pi \hbar^2}\tau(k_F)E_F, \\
\label{conc2}
&\approx & \sigma_0 + q^2 \frac{\hbar v_0^2}{2 V_0^2}n, \quad 2 \pi R^2n\ll 1
\end{eqnarray}
where the carrier (electron) concentration is given by $n=k_F^2/\pi$
with $k_F=E_F/\hbar v_0$ being the Fermi wave vector.
Thus, the low-temperature conductivity at low doping increases linearly with
carrier concentration, in accordance with the experiments. Deviations from
linear dependency can be described taking into account $k$-dependence of the
relaxation time given by Eq.~(\ref{tau}).
We emphasise, that in contrast with \cite{cond-mat2007tan} we deal with
two fitting parameters $R$ and $V_0$ which can be 
deduced using Eq.~(\ref{conc2}) and the experimental data \cite{Nature2005novoselov,cond-mat2007tan,cond-mat2007chen}.
Indeed, from Eq.~(\ref{conc2}) we can define the electron mobility
as $\mu=q \hbar v_0^2/(2V_0^2)$ which {\em does not depend on the screening radius $R$}.
The effective scattering potential then reads
$V_0=v_0\sqrt{q\hbar/(2\mu)}$, and
for the most common samples with $\mu$ ranged from $\sim 10^3$ to  $2\cdot 10^4\,\mathrm{cm^2/(Vs)}$
we have $V_0$ covering the range from $0.1 \,\mathrm{meV}$ to $0.06 \,\mathrm{eV}$.
The screening radius can be estimated from Eq.~(\ref{cond0}) assuming
that $\sigma_0$ is of the order of $4q^2/h$.
Then we have $R$ of the order of $\hbar v_0/ V_0$ covering the range
from $10^{-3}\, \mathrm{cm}$ (high mobility samples) to $10^{-6}\, \mathrm{cm}$ (low mobility samples).
Now one can see that the linear approximation (\ref{conc2}) in terms of $2\pi R^2 n$ 
at $n\sim 10^{12}\,\mathrm{cm}^{-2}$ holds only in relatively low mobility samples,
in accordance with the experimental report \cite{cond-mat2007tan,cond-mat2007chen}.
The difference between $\sigma(n)$ and its linear approximation
can be seen in Fig.~\ref{fig1}.

To consider the low mobility samples properly, we should take into account
the terms proportional to $1/\alpha$ in the solution (\ref{f11}--\ref{f12}).
Note, above all, that the quasiclassical approach is doubtful
at $\alpha\leqslant 1$ since $E_F\tau(k_F)\leqslant \hbar$,
and the carrier mean free path becomes comparable with
its de Broglie wavelength.
Nevertheless, an asymptotic dependence of the solution close to
$\alpha\sim 1$ can give us a clue to what happens in this regime.
The direct integration of Eq.~(\ref{current})
leads to the logarithmic divergence
due to the terms proportional to $f^0_{E_{k +}} - f^0_{E_{k -}}$ in Eqs.~(\ref{f11}--\ref{f12}).
This problem has been solved in Ref.~\cite{cond-mat2007auslender}
introducing some ultraviolet cut off energy $E_c$ which
allows to get a finite value for the conductivity
but still has an unclear physical meaning itself.
In our opinion the divergence of the integral in Eq.~(\ref{current})
is a clear manifestation of obvious limitations inherent in the quasiclassical
approach at $E_F\tau(k_F)\sim \hbar$.
Thus, we expect the ultraviolet cut off $E_c$ to have a quantum mechanical origin.
At $E_F=0$ and zero temperature the quasiparticle energy will fluctuate around
$E=0$ with a variance $\Delta E$ related to the relaxation time via the
uncertainty relation $\Delta E\tau\sim\hbar$.
Obviously, $E_c$ is the maximum energy uncertainty
consistent with two subsequent scattering events (``measurements'')
separated by a time interval $\tau(k_F)$, i. e. $E_c\sim\hbar\tau^{-1}(k_F)$.
Then Eq.~(\ref{current}) can be integrated easily,
and the minimum conductivity for low mobility samples reads
\begin{equation}
\label{cmin}
\sigma_\mathrm{min}=\sigma_0 \left[1+\frac{2}{\alpha}\left(1-\frac{1}{2} \ln \left\vert\frac{4}{\alpha}\right\vert\right)\right].
\end{equation}
The difference between $\sigma_\mathrm{min}$ and $\sigma_0$ given by Eq.~(\ref{cond0})
is shown in Fig.~\ref{fig1} (inset).
We emphasise that Eq.~(\ref{cmin}) could not be well mathematically grounded
in the framework of our quasiclassical model because of quantum effects
which obviously contribute to the conductivity minimum at $\alpha\sim 1$.
What is certainly true is that the additional term in the conductivity minimum stemming from the electron-hole coherence
{\em increases} in low mobility samples and partly compensates the diminution of the leading term given by Eq.~(\ref{cond0}).
This mechanism might be responsible for the non-monotonic dependence of the conductivity minimum
on the impurity concentration (inverse mobility) observed recently \cite{cond-mat2007chen}.

We have also studied the conductivity choosing potential profiles different from one given by Eq.~(\ref{coulomb}).
To give an example for hard-wall potential
when $V(\mathbf{r})=U_0$ at $r\leq r_0$, and $V(\mathbf{r})=0$ at $r > r_0$,
we obtain for the conductivity the same formulas as before
besides the substitutions  $V_0\to \pi r_0 \sqrt{N} U_0$ and $R\to r_0/2$.
The mobility becomes $r_0$-dependent in this case that makes it difficult
to fit our model to the experimental data \cite{cond-mat2007tan}.
Most interesting is, however,
$\delta$-function shaped scattering potential
which can be used as a model for neutral impurities or lattice
imperfections. In this case the Boltzmann equation (\ref{master2}) can be solved easily, and
the conductivity does not depend on the carrier concentration at all,
what, in turn, contradicts the measurements.
Thus, our short-range potential choice given by Eq.~(\ref{coulomb}) fits best to nowadays experimental data.

{\em Thermal conductivity.}
Following our method it is possible to show
that the {\em thermal} conductivity of graphene also has a minimal value
which does not depend on the concentration, but
depends on the temperature according to the Wiedemann--Franz law
as was pointed out in Refs.~\cite{PRB2005gusynin,cond-mat2007dora}.
Indeed, in presence of the temperature gradient $\nabla T$  we have  at $\alpha\gg 1$
in the linear response regime
$f^1_\kappa=(\nabla T/T)\mathbf{v}_\kappa \tau(k) (E_{k \kappa}-E_F)
\left[-\partial f^0(E_{k \kappa})/\partial E_{k \kappa} \right]$.
The further calculations of the thermal flow are very similar to that for
the electrical current.
In particular, the minimum thermal conductivity takes
the form
\begin{equation}
\label{tcond0}
\sigma^\mathrm{th}(E_F=0)\equiv\sigma^\mathrm{th}_0=
\frac{\pi \hbar v_0^2}{12 R^2 V_0^2}T.
\end{equation}
The influence of electron-hole coherence on $\sigma^\mathrm{th}_0$
can be described by Eq.~(\ref{cmin}) in full analogy with the electrical conductivity.

{\em Conclusions.}
We have solved the quasiclassical kinetic equation for carriers
in a single graphene sheet including the off-diagonal elements of the 
distribution function
in the helicity basis.
The analytical solution allows us to investigate the influence of the 
electron-hole coherence on
the minimum conductivity phenomena as well as to discover the limitations
of previous studies based on Boltzmann equation.
We have introduced a special parameter $\alpha$ distinguishing 
the electron-hole coherent and incoherent regimes.
It is noteworthy that $\alpha$
can be deduced directly from the minimum conductivity measurements
since it is incorporated into $\sigma_0$ in a simple way given by
Eqs.~(\ref{cond0}), (\ref{cmin}).
Moreover, our approach successfully describes
the linear dependence of the conductivity above its minimum which is 
usual for low mobility
samples. Finally, we predicted the existence of the thermal conductivity
minimum which was not observed so far.

{\em Acknowledgements.}
We thank Daniel Huertas-Hernando, Inanc Adagideli, Shaffique Adam and Kostya Novoselov for fruitful and
stimulating discussions. This work was financially supported by SFB 689.

\bibliography{graphene.bib}

\end{document}